\documentclass[sigconf,natbib]{acmart}
\usepackage{booktabs} % For formal tables
\usepackage[skip=0pt]{caption}
\usepackage{multirow}
\usepackage{bm}
\usepackage{balance}
\usepackage{algorithm}
\usepackage{algorithmic}

\usepackage[inline]{enumitem}
\usepackage{acronym}

\copyrightyear{2023}
\acmYear{2023}
\setcopyright{acmcopyright}\acmConference[WSDM '23]{Proceedings of the SIXTEEN ACM International Conference on Web Search and Data Mining}{February 27--March 3, 2022}{Singapore}
\acmBooktitle{Proceedings of the Fifteenth ACM International Conference on Web Search and Data Mining (WSDM '23), February 27--March 3, 2022, Singapore}
\acmPrice{15.00}
\acmDOI{10.1145/3488560.xxxxx}
\acmISBN{978-1-4503-9132-0/22/02}


\AtBeginDocument{%
  \providecommand\BibTeX{{%
    \normalfont B\kern-0.5em{\scshape i\kern-0.25em b}\kern-0.8em\TeX}}}

\acrodef{IR}{information retrieval}
\acrodef{NLP}{natural language processing}
\acrodef{MLP}{multi-layer perceptron}
\acrodef{GCN}{Graph Convolutional Network}
\acrodef{PRF}{pseudo relevance feedback}

\allowdisplaybreaks

\author{
	Xiangsheng Li, Xiaoshu Chen, Kunliang Wei, Bin Hu, Lei Jiang, Zeqian Huang and Zhanhui Kang}
\affiliation{
	Tencent Machine Learning Platform Search, \\
	Shenzhen, Guangdong, China
}
\email{lixsh6@gmail.com}

\def\authors{Xiangsheng Li, Xiaoshu Chen, Kunliang Wei, Bin Hu, Lei Jiang, Zeqian Huang and Zhanhui Kang}
\usepackage[T1]{fontenc}
\usepackage{aecompl}
\title{Pretraining De-Biased Language Model with Large-scale Click Logs for Document Ranking}

\begin{document}
	\fancyhead{}
	\begin{abstract}
		Pre-trained language models have achieved great success in various large-scale information retrieval tasks. However, most of pretraining tasks are based on counterfeit retrieval data where the query produced by the tailored rule is assumed as the user's issued query on the given document or passage. Therefore, we explore to use large-scale click logs to pretrain a language model instead of replying on the simulated queries. Specifically, we propose to use user behavior features to pretrain a debiased language model for document ranking. Extensive experiments on Baidu desensitization click logs validate the effectiveness of our method. Our team on WSDM Cup 2023 Pre-training for Web Search won $the$ $1st$ $place$ with a Discounted Cumulative Gain @ 10 (DCG@10) score of 12.16525 on the final leaderboard. 		
 	\end{abstract}

	\maketitle
	%\blfootnote{*Corresponding author.}
	\acresetall
	\section{Introduction}
	Recent advances have shown that pre-trained language models (PTMs) such as BERT~\cite{devlin2018bert}, T5~\cite{raffel2020exploring}, GPT~\cite{radford2018improving} can capture rich semantic information of text and achieve state-of-the-art performance on variance information retrieval tasks~\cite{qiao2019understanding,padaki2020rethinking,li2022cooperative}. However, the pretraining objectives of various PTMs are only based on classical NLP targets~\cite{devlin2018bert} (e.g., Masked Language Modeling and Next Sentence Prediction) and are not carefully explored to better adapt the downstream IR tasks. To address this problem, different pre-training methods with tailored IR objectives are proposed to obtain a better pre-trained language model for downstream IR finetuning tasks. Ma et al.~\cite{ma2021prop} proposed representative words prediction (ROP) task by assuming the sample word set with a higher query likelihood is more ``representative''  to the document. Besides, the dependencies between the inner structures in Wikipedia pages are also exploited to design pretraining tasks~\cite{ma2021pre, wu2022pre} for IR and achieve remarkable retrieval performance compared to the traditional pre-trained language models. These experimental results strongly suggest that traditional PTMs are usually data-hungry on IR tasks, pre-training with suitable IR tasks can effectively boost the performance of IR tasks even on few-shot or zero-shot scenarios~\cite{chen2022axiomatically}.
	
	Despite the success of various IR-based pretraining objectives on PTMs, we observe that these objectives are mostly designed based on manually tailored counterfeit retrieval data where the produced query are assumed as the user's issued query on the given document or passage. As large-scale click logs can be obtained costlessly, we argue that click logs can also be a good resource to pretrain IR-based PTMs. Since the queries are all from the submissions of real users and more close to the query distribution in the downstream IR tasks, designing IR-based pretraining objectives based on click logs provide great potential to improve the downstream IR tasks.
	
	In this work, we explore to use large-scale click logs to pretrain a language model with IR-based objectives. Specifically, we design a CTR prediction task and debiased CTR prediction task as our IR-based pretraining objectives. Furthermore, we extract other sparse features (e.g., BM25, document length, query frequency) and feed them into an ensemble learning model to rerank the candidate list.  Our team on WSDM Cup 2023 Pre-training for Web Search won $the$ $1st$ $place$ with a Discounted Cumulative Gain
	@ 10 (DCG@10) score of 12.16525 on the final leaderboard\footnote{\url{https://aistudio.baidu.com/aistudio/competition/detail/536/0/leaderboard}}. 		
	\section{METHODOLOGY}
	In this section, we present our pipeline which obtains the best ranking performance on the final leaderboard, including four steps: 1) Pre-training with CTR prediction loss; 2) Debiased Pre-training with user behaviors features; 3) Finetuning with pairwise ranking loss; 4) Extracting learning to rank features for each query-document pair; 5) Ensemble learning with all features. We use BERT~\cite{devlin2018bert} as our backbone and feed the concatenation of query, title and content into BERT-based reranker to predict the relevance score.
	
	\subsection{Pre-training with CTR prediction loss}
	\label{section:pretrain}
	Click logs contain rich user behavior information which provides potential relevant query-document pairs. These pairs can be employed to build IR-based pre-training objectives. We do not use the officially provided pointwise CTR prediction loss as we found it will magnify click bias and lead to weak ranking performance. Instead, we use a groupwise CTR prediction loss where the relevance of a clicked document is expected to be higher than other non-clicked documents. The loss is designed as follows:
	
	\begin{equation}
			\label{groupwise_ctr_loss}
		\mathcal{L}_{ctr} = - \log \frac{\exp^{s(q, d^+)}}{\exp^{s(q, d^+)} + \sum_i^{K-1}\exp^{s(q, d_i^-)}}
	\end{equation}  
	
	where $K$ is the group size and a group contains exactly one clicked document and $K-1$ non-clicked documents. Since not all documents in the candidate list are used during pre-training, it can reduce the click bias and yield a better ranking performance compared to pointwise CTR prediction loss.
	Besides, we also use a masked
	language modeling (MLM) loss as our pre-training objective. 
	\begin{equation}
		\label{mlm_loss}
		\mathcal{L}_{mlm} = - \sum_{t_i \in M(x) } \log p(t_i | x \backslash M(x), \Theta)
	\end{equation}  
	where $M(x)$ is the masked tokens of a given sentence $x$ and $x \backslash M(x)$ is the rest tokens. $\Theta$ denotes the parameters of the language model. In particular, we use whole word masking strategy by the provided \textit{unigram\_dict} instead of single word masking. It is shown better performance in various Chinese NLP tasks~\cite{cui2021pre}.	
	
	The final pre-training objective is constructed as follows:
		\begin{equation}
		\label{pre_loss}
		\mathcal{L}_{pretrain} = \mathcal{L}_{mlm} + \mathcal{L}_{ctr}
	\end{equation} 
	 
	\subsection{Debiased Pre-training with user behaviors features}
	To better reduce the impact of click bias during pre-training, we exploit other behavior features to build the IR-based pre-training objectives. Specifically, we use dwell time to filter pre-training group in Section~\ref{section:pretrain}, where the dwell time of clicked document should be longer than other non-clicked documents with a given threshold $\epsilon$. The groupwise ctr prediction loss is 
		\begin{equation}
		\label{debias_groupwise_ctr_loss}
		\mathcal{L}_{ctr} = - \log \frac{\exp^{s(q, d^+)}}{\exp^{s(q, d^+)} + \sum_i^{K-1}\exp^{s(q, d_i^-)}} , t_d(d^+) - t_d(d^-) > \epsilon
	\end{equation}  

	where $t_d$ is the dwell time of a document. In this way, the training samples are with better quality and the clicked document is more confident to be the positive sample. Note, the initialized checkpoint of this task is from Equation~\ref{pre_loss}. 
	
	\subsection{Finetuning with margin ranking loss}
	After pre-training the language model with IR-based objectives, we finetune our model with manually annotated dataset, where each candidate document is marked with a five-level relevance. We employ margin ranking loss to finetune our model. 
	\begin{equation}
			\mathcal{L}_{finetune}= max(0, margin - p(d^+|q) + p(d^-|q))
	\end{equation}

	where $d^+$ is sampled from the documents with relevance higher than or equal to 2 and $d^-$ is sampled from the documents with relevance lower than that of $d^+$. $margin$ is set as 1 in our work.
	
	\subsection{Learning to rank features}
	In our work, we also extract other learning to rank features from the query-document pairs as the learning to rank features, as shown in Table~\ref{sparse_feat}. Specifically, we take the title and content as the whole document. All document features are computed based on title and content.
	
	\begin{table}
		\centering
		\caption{Learning to rank features}
		\begin{tabular}{cc}
			\toprule
			\textbf{ID} & \textbf{feature}                                  \\ \midrule
			1  & query length                             \\
			2  & document length                          \\
			3  & query frequency                          \\
			4  & number of hit words of query in document \\
			5  & BM25 score                               \\
			6  & TF-IDF score                   \\ \bottomrule         
		\end{tabular}
		\label{sparse_feat}
	\end{table}

	\subsection{Ensemble Learning}
	With the learning to rank features and the predicted scores of BERT-based reranker, we feed them into LambdaMart to ensemble the ability of different models. LambdaMart is a state-of-the-art supervised ranker that won the Yahoo! Learning to Rank Challenge (2010)~\cite{wu2010adapting}. We finally aggregate six learning to rank features in Table~\ref{sparse_feat} as well as the predicted scores from BERT-based rerankers in Table~\ref{model_feat}. We first use cross validation to determine the parameters of LambdaMart and then train LambdaMart on the whole validation set. The detailed procedure is as follows:
	\begin{enumerate}
		\item \textbf{Cross validation}: We use Five-fold cross validation to determine the parameters of LambdaMart. Besides, we choose models based on cross validation and finally exclude model 5 since it does not improve ranking performance.
		\item \textbf{Train and inference}: With the determined parameters of LambdaMart and selected rerankers, we train LambdaMart on the whole whole validation set and then calculate the relevance scores in the test set.
	\end{enumerate}

	\begin{table*}[!t]
		\centering
		\caption{Experimental results under different settings. All models except Model 5 are included in the LambdaMart since model 5 does not improve ranking performance in the cross validation.}
		\begin{tabular}{cccccc}
			\toprule
			\textbf{Model ID} & \textbf{Method}           & \textbf{Backbone} & \textbf{Pretrain step} & \textbf{Finetune step} & \textbf{Submission DCG} \\ \midrule
			1                 & CTR pre-training          & BERT-24           & 1700K                & 5130                   & 11.96214                \\
			2                 & CTR pre-training          & BERT-24           & 1700K                & 4180                   & unk                     \\
			3                 & CTR pre-training          & BERT-12           & 2150K                & 5130                   & 11.32363                \\
			4                 & CTR pre-training          & BERT-24           & 590K                 & 5130                   & 11.94845                \\
			5\rlap{$^*$}                 & CTR pre-training          & BERT-24           & 1700K                & 4180                   & unk                     \\
			6                 & Debiased CTR pre-training & BERT-24           & 1940K                & 5130                   & unk                     \\ \bottomrule         
		\end{tabular}
		\label{model_feat}
	\end{table*}

	\section{Experiment}
	\subsection{Experimental settings}
	In our work, we select BERT-base (12 layers) and BERT-large (24 layers) as our backbone and use a linear layer to predict the relevance score. Masking ratio in Equation~\ref{mlm_loss} is set as 0.15. $\epsilon$ in Equation~\ref{debias_groupwise_ctr_loss} is set as 8 seconds. The parameters of LambdaMart after cross validation are: 300 training epochs, 100 leaves, 0.05 learning rate. We released our code implemented by Pytorch and PaddlePaddle at \url{https://github.com/lixsh6/Tencent_wsdm_cup2023}. 
	
	\subsection{Results}
	We list experimental results under different settings in Table~\ref{model_feat}. $unk$ denotes that we did not submit it to the leaderboard. We can observe that using BERT-large (24 layers) can achieve better ranking performance compared to BERT-base (12 layers). In particular, debiased CTR pre-training (Model 6) can achieve better performance than CTR pre-training in our cross validation experiments, as shown in the next subsection. Since we did not submit it to the leaderboard, we analyze it by visualizing feature importance of LambdaMart. Finally, we feed 6 six learning to rank feature in Table~\ref{sparse_feat} and five BERT-based prediction scores in Table~\ref{model_feat} (exclude Model 5) into LambdaMart, the DCG@10 on leaderboard achieves 12.16525. 
	
	\subsection{Feature importance}
	\begin{figure}[!t]
		\centering
		\includegraphics[width=80mm,height=54mm]{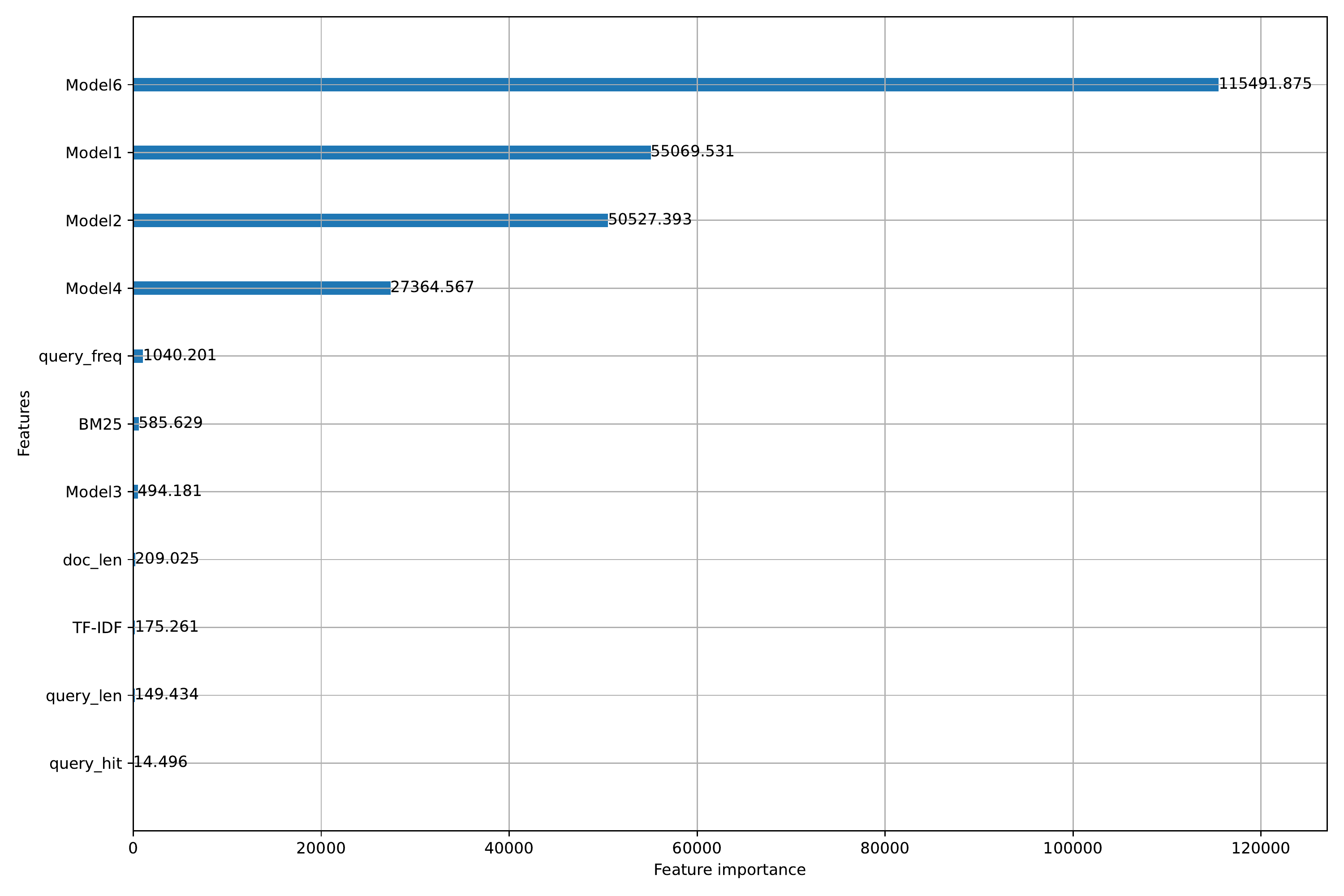}
		\caption{Feature importance in LambdaMart.}
		\label{feat_importance}
	\end{figure}

	We visualize the feature importance of LambdaMart in Figure~\ref{feat_importance}. We find the prediction scores of most BERT-based rerankers cover most of weights in LambdaMart, which suggests that pretrained language model is an important reranker compared to traditional IR methods (e.g., BM25, TF-IDF). In addition, we notice that model 6 plays the most important role in LambdaMart with about twice weights of the second rank. It illustrates that debiased CTR pre-training can effectively boost ranking performance compared to the traditional CTR pre-training. 
	 
	\section{Conclusion}
	In this paper, we introduce our method on WSDM Cup 2023 Pre-training for Web Search which won $the$ $1st$ $place$ with a Discounted Cumulative Gain
	@ 10 (DCG@10) score of 12.16525 on the final leaderboard. We have the following conclusions:
	\begin{enumerate}
		\item Pre-training with groupwise CTR prediction loss leads to a better ranking performance in the downstream task compared to pointwise CTR prediction loss. It is due to the high click bias if modeling on the full document list.
		\item Whole word masking can effectively boost the ranking performance.
		\item Debiased Pre-training with user behaviors features can effectively reduce the click bias in the click logs, leading to a better pretrained language model.
		\item Using BERT-large reranker can achieve better ranking performance than BERT-base reranker.
	\end{enumerate}

	Besides, we also attempt other popular pre-training strategies such as retrieval-oriented pretraining with decoders~\cite{liu2022retromae}, T5 as reranker~\cite{zhuang2022rankt5}, etc. But we do not find their effectiveness in this competition task. We believe click logs can be a valuable resource to pretrain an IR-based language model and look forward to studying more this area in future work.
	%\newpage
	\bibliographystyle{ACM-Reference-Format}
	\balance
	\bibliography{reference}
\end{document}